\documentclass[12pt]{revtex4-1}%{iopart}
\usepackage{revsymb4-1}
\usepackage{graphicx}

\begin{document}

\title{Spatially-resolved potential measurement with ion crystals}
\author{M.~Brownnutt$^1$, M.~Harlander$^{1}$, W.~H\"{a}nsel$^{1,2}$ and R.~Blatt$^{1,2}$ }

\address{$^1$ Institut f\"{u}r Experimentalphysik,
Universit\"{a}t Innsbruck,
Technikerstrasse 25,
A-6020 Innsbruck, Austria}

\address{$^2$ Institut f\"{u}r Quantenoptik und Quanteninformation
der \"{O}sterreichischen Akademie der Wissenschaften,
Technikerstrasse 21a,
A-6020 Innsbruck, Austria}

\date{\today}

\begin{abstract}
We present a method to measure potentials over an extended region using one-dimensional ion crystals 
in a radio frequency (RF) ion trap.
The equilibrium spacings of the ions within the crystal allow the determination of the external forces acting at each point.
From this the overall potential, and also potentials due to specific trap features, are calculated.
The method can be used to probe potentials near proximal objects in real time, and can be generalized to higher dimensions.
\end{abstract}

% insert suggested PACS numbers in braces on next line
% insert suggested keywords - APS authors don't need to do this
%\keywords{Pick some %single-ion probe, ion trap, photo emission, patch charges, electric-field measurement}
%\submitto{APL}
\maketitle

Trapped ions stand as one of the pre-eminent realizations for scalable quantum computation \cite{ARDA:2004,Haeffner:2008}.
One prominent method of scaling such systems to many qubits
requires the shuttling of ions through structured geometries \cite{Kielpinski:2002}.
While such structures and shuttling processes can be simulated, stray static charges or minor misalignments in fabrication
can cause the actual shuttling behavior to differ markedly from that calculated \cite{Hensinger:2006}.
Furthermore, the operation of nominally identical traps can vary significantly \cite{Leibrandt:2009}.
The direct measurement of the electric fields in an ion trap thus provides an invaluable tool in understanding and optimizing trap operation.
Fortunately trapped ions themselves can serve as immensely sensitive probes of their environment.
They have been used to infer the spectral intensity of electric field noise \cite{Deslauriers:2004},
 as well as measuring AC \cite{Biercuk:2010} and DC \cite{Harlander:2010} forces and electric fields
equivalent to a few elementary charges at a distance of a millimetre.
This paper describes the use of an extended, one-dimensional (1D) string of ions
to measure the electrical potentials due to ion trap features.
The method is simple and robust, requiring neither resolved-sideband techniques 
nor any assumptions of the trap's harmonicity \cite{Huber:2010}.
It can measure both the total effective potential along a line,
and the individual contributions due to selected electrodes.
The technique can also be extended to investigate external objects,
such as the charging of nearby surfaces \cite{Harlander:2010, Maiwald:2009}.

We consider a 1D string of singly-charged ions in a confining potential.
The individual ions are acted upon by both the confining potential and the Coulomb repulsion of the other ions.
Each ion, $i$, at position, $x_i$, experiences a Coulomb force due to all other ions, $j$, given by

\begin{equation}
F_{\mathrm{ion}}^{(i)}=\frac{e^2}{4\pi\epsilon_0}\sum\limits_{j\neq i} \frac{|x_i-x_j|}{( x_i-x_j)^3}.
\label{eq:IonForce}
\end{equation}

\noindent At equilibrium this force is equal and opposite to
the force due to the external confining potential, $F_{\mathrm{ext}}(x_i)$.
Using the ion positions as interpolation points
the function, $F_\mathrm{ext}(x)$, can be numerically integrated to give the instantaneous confining potential in 1D
(except for an unknown integration constant).

In an ion trap, particularly a segmented ion trap \cite{Schulz:2006},
the total potential is due to the combined effects of a number of different DC electrodes,
plus a (small) contribution from any axial component of the RF  pseudopotential.
In addition to this combined information,
it is often useful to isolate the effects due to one particular electrode or feature.
This can be achieved by taking repeated measurements,
each time varying the voltage on the electrode of interest.
The potential, $\psi(x)$, will depend on the voltage on the electrode of interest, $V_\mathrm{A}$,
and also on all other voltages, applied to all other electrodes.
As the latter are held constant, they simply provide a constant background offset
and are collectively termed $V_\mathrm{B}$. By simple use of the superposition principle,
the difference between two potential measurements
gives the effect due to the electrode of interest:

\begin{equation}
\left[\psi(x,V_\mathrm{A},V_\mathrm{B})-\psi(x,V_\mathrm{A}-\delta,V_\mathrm{B})\right]/\delta=\psi(x,V_\mathrm{A}=1,V_\mathrm{B}=0).
\label{eq:ElectrodeEffect}
\end{equation}

\noindent The addition of a voltage to one electrode will necessarily move the positions of the ions, $x_i$.
However, by interpolation, the extended nature of the string means that the potential $\psi(x,1,0)$
can be calculated for all $x$ where the two data sets,
$\psi(x,V_\mathrm{A},V_\mathrm{B}), \psi(x,V_\mathrm{A}+\delta,V_\mathrm{B})$, overlap.
Uncertainty due to numerical errors can be significantly decreased by
averaging over the results of potentials for many values of $\delta$.
The fact that the ion string moves as $\delta$ is varied also allows even larger regions to be explored.

The apparatus used to implement such measurements is described elsewhere \cite{Splatt:2009},
and reviewed here. $^{40}$Ca$^+$ ions are trapped in a segmented, surface, Paul (RF) trap.
An RF potential of V$_0$ = 300\,V (0-peak) and frequency $\Omega_{\rm{T}}$ = 2$\pi\times$10.125\,MHz
provides radial confinement with motional frequencies $\omega_y$, $\omega_z$ = 2$\pi\times$(250, 800)\,kHz.
The axially confining potential is provided by voltages (-20\,V $< V <$ 60\,V)
applied to five pairs of DC segments,
and has a curvature much less than that of the radial potential.

The ion string is Doppler cooled using 397\,nm and 866\,nm laser light,
parallel to the trap surface.
The 397\,nm light scattered by the ions is imaged using a custom-made lens
(f\# = 1.7, NA = 0.28, focal length = 67\,mm) onto a CCD camera to
provide a resolution of 2\,$\mu$m per pixel.
Cooling to the Doppler limit is not required:
the individual ions must simply be resolvable on the CCD image.
The ions' micromotion is minimized in the $y$-direction (parallel to the trap) by ensuring
that the position of the ions observed on the camera does not vary as a function of RF power \cite{Berkeland:1998}.
Coarse micromotion reduction in the $z$-direction (normal to the trap) is achieved by adjusting
the height of the ions above the trap until the images of the individual ions are well localized on the CCD camera.

\begin{figure}
\includegraphics[]{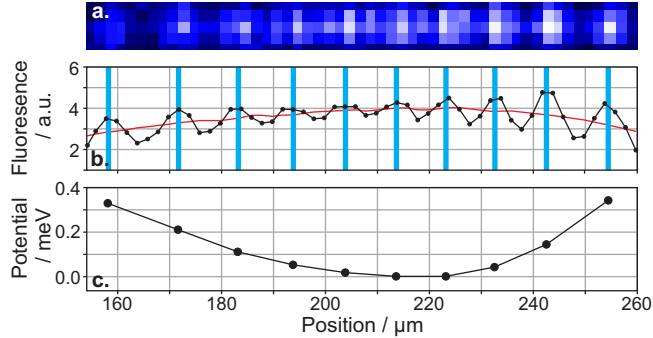}
\caption{\label{Fig1}
Calculation of a 1D potential.
(a) The ions are imaged, typically with a 100\,ms exposure time.
%(peak counts of \sym\{\}\,per pixel).
(b) The counts for each column are summed (dots).
To compensate for non-uniform illumination and imaging artifacts a
slowly varying background (red line) is subtracted.
The ion positions (blue lines) are then determined using a weighted fit.
There are a number of more involved algorithms for identifying ion positions \cite{Eble:2010}, 
though the simple method used here was sufficient to give a 
positional uncertainty of less than $\pm$0.5\,$\mu$m, or 0.25\,pixels.
(c) The axial potential is reconstructed by numerical integration
of the forces calculated by Eq.\,\ref{eq:IonForce}.
The offset of the potential has been arbitrarily chosen such that the minimum of the potential is zero.}
\end{figure}

\begin{figure}
\includegraphics[]{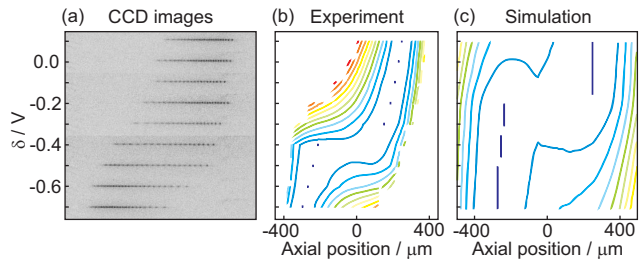}
\caption{\label{Fig2}
Analyzing a shuttling routine.
The initial axial potential was created by applying
($V_1$,$V_2$,$V_3$,$V_4$,$V_5$) = (40.5, 4.64, 30.8, 4.50, 40.5)\,V
% These data are from 2008-01-14. LB 2 P 171-172
to the 5 DC electrode segments. It was subsequently varied by increasing $V_2$
and decreasing $V_4$ by equal amounts, $\delta$.
At various points ions were lost and/or reloaded,
though this did not detrimentally affect the measurement.
For each measurement the string consisted of between 18 and 25 ions.
(a) CCD images of the ion strings for different values of $\delta$.
(b) A contour plot of the measured potential as a function of $\delta$.
Equipotential lines are spaced by 0.4\,meV.
(c) Simulation data for similar voltages to those used in the experiment (within 1\,V).
This shows that the experimentally observed behaviour can be qualitatively reproduced in simulations.
When the exact voltages used are simulated
the ``hump'' observed at $\delta$ = -0.2\,V is so large that a
double-well potential is formed,
and the ions are trapped - and remain - in one of the two wells.}
\end{figure}

The analysis of a typical set of data is shown in Fig.\,\ref{Fig1}.
The ions are imaged (a) and their positions calculated from a fit to the detected fluorescence (b).
From these positions the potential is reconstructed (c).
Having evaluated a single potential,
variation of this potential - for processes such as shuttling - can also be analyzed.
Fig.\,\ref{Fig2} shows a series of ion-string images,
each taken for different sets of electrode voltages.
These were used to calculate the potential seen by the ions
as they were moved from left to right along the trap.
Use of only a few ions would provide information about the well minima.
However, the presence of a large number of ions allows visualization of the entire potential in an instant.

The exact shape of the confining potential (of order mV) is due to small
differences between large electrode voltages (of order V).
Any imprecision in a computer simulation of the trap
can give rise to qualitatively different predicted potentials.
For example, for the scenario in Fig.\,\ref{Fig2}
the simulation was qualitatively different
(no shuttling was observed between the two separate wells).
The direct measurement using ions is clearly a more sensitive
and accurate method.

\begin{figure}
\includegraphics[]{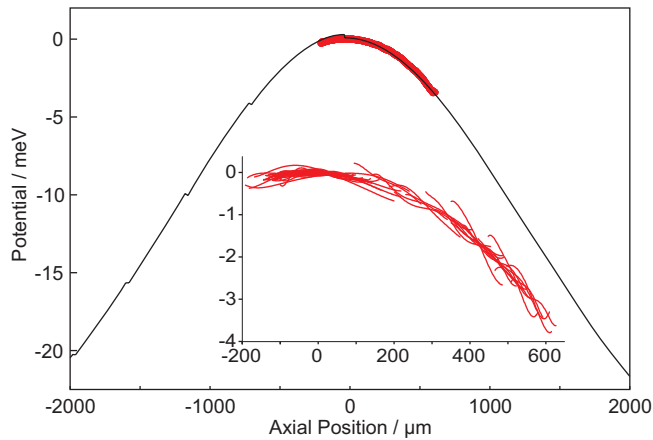}
\caption{\label{Fig3}
Characterization of the potential due to a given feature.
The thin solid line shows the potential
along the trap axis due to a voltage $V_3=1$\,V applied to the middle DC electrode,
as calculated by boundary-element-method simulations.
The discontinuities are an artefact of the 10\,$\mu$m grid resolution of the simulation output.
The bold line is the result from ion-string measurements.
Potentials were measured for many different values of $\delta$.
The pairwise differences between these measurements were used to calculate the effect of
the central electrode for the region in which the measurements overlapped
(\textit{c.f.} Eq.\,\ref{eq:ElectrodeEffect}).
Results of each pairwise measurement are shown in the insert.
These results were then averaged to give the line in the main figure.}
\end{figure}

Finally, the effect of a single electrode can be ascertained.
Typical results are shown in Fig.\,\ref{Fig3}.
Measurements were taken of the total trap potential,
with different voltages, $V_3+\delta$, applied to the middle DC electrode.
The effect of this single electrode was calculated from Eq.\,\ref{eq:ElectrodeEffect}.
Although the ion string was typically only 200\,$\mu$m long,
the potential due to the single electrode could be measured over an extended range ($\sim$1\,mm),
as the ion string was moved to different positions.

In addition to measuring the effects of features on the ion trap itself, the
method presented could be used to measure electrical effects on other objects.
Short strings of ions have previously been used as real-time field probes \cite{Harlander:2010}.
The use of extended ion strings for such applications
would allow spatial resolution of the effects under consideration.

The method described here is one dimensional.
In the limit of strong radial confinement
it can be assumed that the height of the ion string above the trap does
not change significantly.
If, however, the radial confinement is weak, the ions' height may change.
They will then sample the potential $\psi(x,z)$.
For applications such as mapping out a 2D potential, this may be of benefit.
The dimensionality of the measurement could also be increased by
using 2D \cite{Block:2000, Buluta:2009} or 3D \cite{Hornekaer:2002} ion crystals.
A generalization of Eq.\,\ref{eq:IonForce} would allow
mapping of potentials in two or (with the addition of suitable optics) three dimensions.

In conclusion, a method has been presented to use extended strings of trapped
ions as a sensitive probe of electrical potentials. The method can be
used both to measure the total potential experienced by the string
and to measure the potential due to a single electrode or feature.
This can find applications for characterizing and optimizing ion trap geometries,
as well as in sensitive measurements of surface physics.

\end{document}